\documentclass[preprint,12pt,authoryear]{elsarticle}

\usepackage{amsmath,amssymb,amsfonts}

\usepackage{array,longtable} 
\usepackage{url}
\date{}

\journal{Journal of Informetrics}

\begin{document}

\begin{frontmatter}

\title{Dynamics of senses of new physics discourse: co-keywords analysis}

\author[hse]{Yurij L. Katchanov\corref{mycorrespondingauthor}}
\cortext[mycorrespondingauthor]{Corresponding author}
\ead{yurij.katchanov@gmail.com}

\address[hse]{Institute for Statistical Studies and Economics of Knowledge, National Research University Higher School of Economics  
20 Myasnitskaya Ulitsa, Moscow 101000, Russian Federation}

\author[aaas]{Yulia V. Markova} 
\ead{yulia.markova@gmail.com}

\address[aaas]{American Association for the Advancement of Science   
1200 New York Ave NW, 20005, Washington, DC, USA}

\begin{abstract}
The paper presents a longitudinal analysis of the evolution of new physics keywords co-occurrence patterns. For that, we explore the documents indexed in the INSPIRE database from 1989 to 2018. Our purpose is to quantify the knowledge structure of the fast-growing subfield of new physics. The development of a novel approach to keywords co-occurrence analysis is the main point of the paper. In contrast to traditional co-keyword network analysis, we investigate structures that unite physics concepts in different documents and bind different documents with the same physics concepts. We consider the structures that reveal relationships among concepts as topological and call them ``physics senses''. Based on the notion of trajectory mutual information, the paper offers clustering of physics senses, determines their period of life, and constructs a classification of senses' ``authority''. 
\end{abstract}

\end{frontmatter}

\section{Introduction}
Throughout the modern history of physics, high energy physics (or HEP for convenience) has been and remains one of the main driving forces of conceptual and methodological innovations. It is also a front-runner in the development of scientific tools for publication and communication. E.g., the INSPIRE\footnote{\url{https://inspirehep.net}} database is the leading publication platform indexing HEP publications \citep{hecker_four_2017}. All these factors make HEP an interesting subject of study inside the science of science approach \citep{Fortunatoeaao0185} (hereafter, abbreviated as SciSci). One of the principal objectives of SciSci is to comprehend the emergence of new knowledge, and HEP can provide a first-choice object of research in this regard.

The somewhat bureaucratic term of ``the Standard Model'' traditionally defines the monumental achievement of the twentieth-century HEP---the product of efforts by thousands of researchers spanning over 50 years \citep{Langacker_2017}. The Standard Model constitutes a contemporary view of particle physics that combines theoretical and experimental results of research into the properties and interactions of particles. With no deviations from the Standard Model discovered to date, it remains on the cutting edge of our experimentally proven understanding of the fundamental laws of the microworld. However, there has to be a new physics beyond the Standard Model---something that researchers have been working to uncover for decades now. This smart term is used to describe the fundamental layer of the physics reality at the core of the structure of the universe, destined to replace the Standard Model as the leading theory.

New physics is not a specific theory, but rather a collective term for everything that is at variance with the Standard Model \citep{Ghosh_2020}. Judging by abundant circumstantial evidence, new physics does exist but remains hard to define. For that reason, a multitude of various methods have been applied in search of its manifestations. Spanning decades, these concentrated and diverse experimental and theoretical quests for new physics make it into a unique testing ground for the development of new methods of the quantitative study of science. 

This paper presents a longitudinal analysis of the evolution of new physics keywords co-occurrence patterns based on the documents indexed in the INSPIRE database from 1989 to 2018. We consider keywords co-occurrence structure as an operationalization of the new physics knowledge structure. These patterns reveal the relationships between new physics knowledge entities. 

Our purpose is to quantify the knowledge structure of the fast-growing new physics' subfield. The issue addressed in the paper is the development of a novel approach to keywords co-occurrence analysis.

The remainder of this paper is organized as follows. Section~2 refers to related studies on a given topic. Section~3 presents the methodological framework. Section~4 explains the data used in our analyses. Section~5 reports our main results and generalizes the findings of the paper. Finally, in Section~6, we draw the conclusions.

\section{Background}
Quantitative research on physics, physicists and the discourse of physics has by now become a well-established tradition \citep{sinatra_century_2015, Colavizza_2016, battiston_taking_2019, Sun_2020}. It allows for a shift from descriptive analysis and general arguments to more keen and detailed research on the evolution of physics as a scientific field \citep{liu_predicting_2019}. The availability of extensive publication databases and advanced data processing tools open the door to new interdisciplinary research opportunities. Take, e.g., the study of the evolution of research interests or physics subject areas \citep{jia_quantifying_2017, aleta_explore_2019, Zeng2019, Coccia_2020}, physics research institutions \citep{Katchanov_2016}, careers pathways \citep{Deville_2014, Petersen_2018, Xing_2019}, or strategies and talent of physicists \citep{Pluchino_2019, Tripodi_2020}. Due to its nature, physics has its own information environment, and researchers have access not only to general commercial databases like Web of Science or Scopus but also to specific physics publications repositories like arXiv, INSPIRE, or APS. Thanks to the openness of these specialized data sources, researchers may use the full range of quantitative methods to study physics.

Quantitative analysis of physics concepts provides a good opportunity to explore the relations between external classifications of publications and the outcomes of community detection algorithms \citep{palchykov_ground_2016}. The authors demonstrated that classifications made by authors had some discrepancies from automatically retrieved communities. These differences may carry important information about relations between physics fields that the external observer does not realize. M.~Krenn and A.~Zeilinger used a large corpus of publications covering about a hundred years to train a neural network and try to predict future trends in physics \citep{krenn_predicting_2020}. A large body of literature opens the way to interesting studies of cognitive aspects of science. Thus, S.~Milojevi{\'{c}} compared the growth of scientific literature and the lexical diversity of research articles titles. She found that periods of fast cognitive growth did not coincide with the rise of publication growth \citep{milojevic_quantifying_2015}. 

Recent advances in text analysis methods have facilitated detailed quantitative research on physics concepts. Thus, M.~Herrera, using publications and PACS categories, built physics fields and traced their evolution for a period of \(20\) years \citep{herrera_mapping_2010}. M.~Fontana et al. \citep{Fontana_2020} analyzed more than \(230\) thousand articles published in \(8\) journals of the American Physical Society and \(2.4\) million citations to measure novelty and interdisciplinarity. R.K.~Pan et al. studied the evolution of interdisciplinary research in physics \citep{pan_evolution_2012}. They constructed a network of publications and showed that interactions between physics fields were increasing, and the area of interdisciplinary research was growing. Using publications data spanning more than a century, M.~Perc explored patterns and trends of physics terminology. The analysis revealed that the evolution of scientific paradigms in physics was governed by principles of self-organization \citep{perc_selforganization_2013}.

Machine learning methods gained popularity in SciSci in recent years \citep{Xu_2020}. Thus, K.~McKeown et al. \citep{mckeown_predicting_2016} presented a system that predicts the impact of a physics concept; M.~Chinazzi et al. \citep{chinazzi_mapping_2019} mapped research space in physics using the American Physical Society publications database; A.~Palmucci et al. \citep{ Palmucci_2020} suggested a representation for the relative movements of physics' domains in a multi-dimensional space based on the PACS codes.

Network analysis has a crucial part to play in the SciSci \citep{ZENG2017} these days. Given its universal scope, network analysis allows the study of various aspects of physics as a scientific field. In recent years, a wealth of research has been produced on co-authorship networks in physics \citep{heiberger_choosing_2016, silva_coauthorship_2019}, co-citation networks \citep{zhang_characterizing_2013, renoust_flows_2017, dearruda_how_2018}, and keyword co-occurrence networks \citep{Palchykov_2021_1, You_2021, Karimi_2021}. In the past two decades, the application of co-keyword analysis, first used by Callon et al. \citep{Callon_1983}, was both far-ranging and far-reaching, covering relevant aspects of research topic networks \citep{Cobo_2011, Yi_2012, Leydesdorff_Nerghes_2017, Radhakrishnan_2017, Wen_2017, Behrouzi_2020, Wang_2021}.

When studying co-authorship networks, co-citation or co-keyword networks, graph theory is often perceived as a mathematical framework underlying the SciSci \citep{Mingers_2015}. At the same time, quantitative science studies are evolving, and so is their mathematical apparatus. In a number of papers, a depiction produced with customary network tools is supplemented with that created with tools from applied topology. E.g., \citep{Salnikov_2018, Christianson_2020} use a simplicial complex approach to word co-occurrences in order to explore the conceptual landscape of mathematical publications. A.~Patania et al. \citep{Patania_2017b} develop simplicial descriptions of publications and persistent homology to track the growth and development of collaborations based on co-authorship data.

\section{Methods}
\subsection{Problem formulation}
We apply the co-word analysis technique in order to map the structural dynamics of the ``new physics'' subfield. This analysis has proven effective for the study of relationships and interrelationships among scientific concepts \citep{vanRaan2019}. The basic principle of co-word analysis is to characterize the underlying texts through the use of co-occurrence networks  consisting of nodes that represent the keywords and edges between them \citep{Callon_Courtial_1991}.

Physics knowledge is a complex system, at its basic level consisting of physics concepts and their interactions. From the SciSci point of view, the complexity of the physics knowledge system largely results from the variety of interactions between concepts. The network approach focuses on the analysis of pairwise relationships between concepts (terms, keywords), invariably reducing multidimensional structures and losing information. Nevertheless, the recent decades saw a great number of intellectual structures described successfully as networks where co-occuring pairs of concepts are connected by links \citep{Li_2016, Wen_2017, Radhakrishnan_2017, Loazano_2019}. These achievements are prompting the quantitative science studies to develop new methods of analysis \citep{Cheng2020, Joslyn_2021}.

The scientometrics community is in agreement that system of well-sepa\-ra\-ted and well-connected clusters representing keywords in a co-occurrence network can ensure an adequate description of a research discourse \citep{vanEck2014, Hosseini_2021}. When such network representation of scientific discourse is assumed, a strong assumption is made: the overall interplay among the concepts is suggested to be completely described by pairwise co-occurrence data. In reality, however, the constraints in a co-occurrence network may prove to be fairly weak, and so the identification of the true structure of research themes may require that more complex interactions are taken into account that help reveal more interplay among physics concepts. The problem of the context in which a keyword co-occurrence network is generated was first described in \citep{Bornmann2018} and was studied in more detail in \citep{Cheng2020}. However, the proposed heuristic solutions concern the use a citation network and are beyond the scope of keyword co-occurrence analysis.

Research communication has a distinguishing feature, which is that all components of the system influence each other in a direct or indirect way. Each scientific publication features several concepts simultaneously, and one should not assume \emph{a priori} that their interactions can be explained in terms of dyads (see details in \citep{Battiston_2020}).

\subsection{Physics sense: definition}
The sense of a physics concept cannot be fully explained by its formal definition. Its understanding is greatly improved by a series of examples illustrating its use in documents. For working physicists, these examples represent both motivation and the sense of a concept. It is for this reason that exploring the co-occurrence of physics concepts can prove to be a significant element in the study of the cognitive structure of physics. The sense helps to reveal the essence of a physics concept in the research context. A physics sense determines the place of the concept in scientific communication. The sense introduces the ``whole---part'' relationship, making the concept essential as part of a whole which is the system of a physics discourse.

Specification of documents that mention a physics concept is imperative for the understanding of its  physics sense. We consider a physics sense in the framework of an ``information field''. The concept of the field emerges naturally as one analyzes the forming of a physics sense within a system of relationships between physics concepts. Information field is a holistic structure that describes both direct and indirect mutual influences of physics concepts in documents. In this paper, the intuitive viewpoint is that an information field that determines a physics sense is a topological space of physics concepts and their properties.

Following the generally accepted definition (see, e.g., \citep{Frege, Deleuze}), one understands physics sense as a proposition including a set of physics concepts. The intellectual value of a concept is the result of its position in a complex network of semantic relations. The study of physics discourse in its entirety involves a structural comparison of the concepts. Our methodology supposes that to study physics concepts, one needs to construct statistical relations between them. This type of relations we call ``sense''.

\begin{figure}[htb!]
\centering
  \includegraphics[scale=0.91]{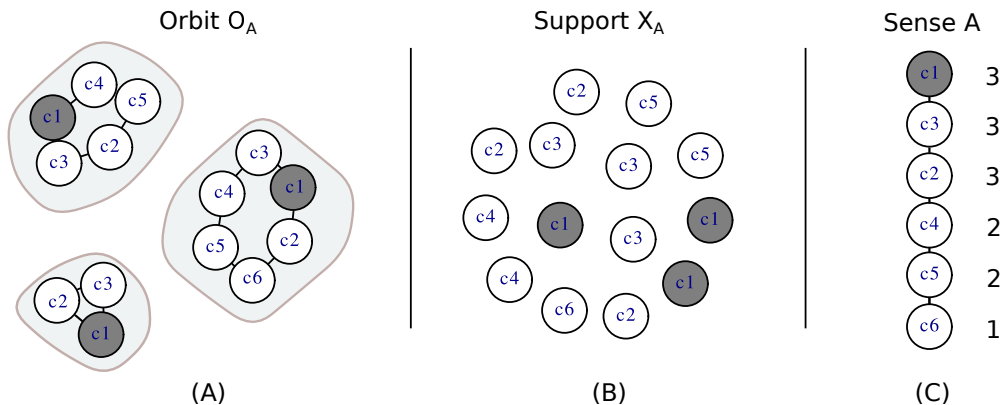}
\caption{Construction of physics sense. Part A shows how the documents are related together in the orbit \(O_{A}\) of the concept \(x_{{\alpha}0}\equiv{c_{1}}\) (marked with a dark circle). Part B depicts the support \(X_{A}\) of a sense \(A\). Part C displays the sense \(A\).}
\label{fig1}      
\end{figure}

For our further reasoning we will need auxiliary concepts. Let us refer to the set of all documents that mention the physics concept \(x_{{\alpha}0}\) as the orbit \(O_{A}\) of that concept. We will understand the spectrum \(S_{A}\) of physics concept \(x_{{\alpha}0}\) as a set of physics concepts \(x_{{\alpha}0},x_{{\alpha}1},\ldots ,x_{{\alpha}k}\) corresponding to the orbit \(O_{A}\). (Clearly, \(x_{{\beta}0}\in S_{A}\Leftrightarrow x_{{\alpha}0}\in S_{B}\), etc.) Let \(X_{A}\) (with or without indices) be a sample of \(M\) physics concepts that occur in the orbit \(O_{A}\) with frequencies \(m_{{\alpha}0},m_{{\alpha}1},\ldots ,m_{{\alpha}k}\), such that \(\sum_{\alpha}m_{\alpha}={M}\).

Let \(\sigma_{A}\) be a configuration of the set \(X_{A} = \lbrace x_{\alpha}\rbrace_{\alpha\in A}\). The configuration \(\sigma_{A}\) can be represented in the form
\begin{equation}
\label{config}
(\forall{x_{\alpha}}\in S_{A}\subset O_{A})\colon \sigma_{A} = \begin{pmatrix} x_{{\alpha}0} & x_{{\alpha}1} & \ldots & x_{{\alpha}k} \\ m_{{\alpha}0} & m_{{\alpha}1} & \ldots & m_{{\alpha}k} \end{pmatrix}{.}
\end{equation}
The set \(X_{A}\) will be called hereafter the support of a physics sense \(A\) (see Fig.~\ref{fig1}). A topological structure on the support \(X_{A}\) is a collection \(\tau_{A}\) of subsets of \(X_{A}\) which includes the empty set \(\emptyset\) and the whole support \(X_{A}\) and which is such that: a) the intersection of any number of elements of \(\tau_{A}\) belongs to \(\tau_{A}\) and b) the union of any set of elements of \(\tau_{A}\) belongs to \(\tau_{A}\) \citep{Singh_2019}. In our case, the collection of all subsets of \(X_{A}\) is a topology on \(X_{A}\), and \(\tau_{A}\) is called the discrete topology.

One can think of \(S_{A}\) as having been obtained by ``identifying'' each pair of equivalent elements of the support \(X_{A}\). Hence, the surjective map \(\pi\colon X_{A}\rightarrow{S_{A}}\) carries each point of the support \(X_{A}\) to the equivalence class \(\omega\) containing it. In the quotient topology induced by \(\pi\), the set \(\Omega\) such that \(\Omega = \cup_{\alpha\in X}\omega_{\alpha}\) is called a quotient space of \(X_{A}\) \citep{Singh_2019}. For further details, we refer to \ref{Framework}.

In this paper, a \emph{physics sense} \(A\) of concept \(x_{{\alpha}0}\) is a finite topological space \((X_{A},\tau_{A})\)---a support \(X_{A}\) with a discrete topological structure \(\tau_{A}\). By putting it this way, we symbolically equate the space \((X_{A},\tau_{A})\) with its focal point---the concept \(x_{{\alpha}0}\). The main point here is to realize that the topological structure \(\tau_{A}\) is the elementary property of physics sense as a whole. It means this structure refers to the physics concepts collection \(X_{A}\), but not to its parts or connections between these parts.

Knowing the configuration \(\sigma_{A}\) Eq.\eqref{config} of the physics sense \(A\) enables us to define a state function \(h(A)\) \(=-\sum{p(\omega)}\ln{p(\omega)} = \sum_{\alpha}p(\alpha)\ln{p(\alpha)}\) (see Eq.~\eqref{eq:g3})---the  topological entropy, where \(p(\alpha)\) is the probability of the realization of the physics concept \(x_{\alpha}\) in the orbit \(O_{A}\). The interpretation of entropy \(h(A)\) is that it is a measure of the number of arrangements the physics concepts could be in.

\subsection{Basic quantities}
\subsubsection{Dimension}
The quantity of dimension  \(\dim (A)\) Eq. \eqref{eq:dim} is the number of the physics concepts in the spectrum \(S_{A}\). It is possible to interpret a dimension  as a number of degrees of freedom of the  information field of a physics sense. A more straightforward interpretation of a dimension is that \(\dim\) constitutes the number of information channels that connect a single sense with others.

\subsubsection{Trajectory mutual information}
Thus, we attempted to describe the sense of a physics concept through a configuration of the information field where the concept functions in the discourse. In order to characterize the thematic structure of physics discourse itself, we need to examine the configuration of the field induced by all relevant physics senses. Since we have conceptualized a physics sense as a topological space of concepts within a single orbit, it is only natural to consider the interaction of physics senses as a topological space constructed on the intersection of respective orbits. If the orbits of physics concepts intersect, it indicates that the concepts share a common context and we can evaluate its significance using the concept of topological entropy.

Let us consider physics senses \(A\) and \(B\), i.e., topological spaces \((X_{A},\tau_{A})\) and \((X_{B},\tau_{B})\) with orbits \(O_{A}\) and \(O_{B}\) and spectra \(S_{A}\) and \(S_{B}\) so that  \(x_{\alpha0}\in S_{B}\) (and, respectively, \(x_{\beta0}\in S_{A}\)). On the intersection of orbits \(O_{A}\) and \(O_{B}\) as well as spectra \(S_{A}\) and \(S_{B}\), a topological space \((X_{AB},\tau_{AB})\) is defined that describes the interplay between the physics senses \(A\) and \(B\). One can easily define the support \(X_{AB} = X_{A}\cap X_{B}\)
\begin{equation}
\label{supp}
\left(O_{A}\cap{O_{B}}\right)\left(\forall{x_{\alpha\beta}}\in{S_{A}\cap{S_{B}}}\right)\colon X_{AB}=\left\lbrace x_{\alpha\beta}\right\rbrace_{\alpha\beta\in{AB}},
\end{equation}
the discrete topology \(\tau_{AB}\) on \(X_{AB}\), and the quotient space \((\Omega_{AB},\varpi_{AB})\) as shown in \ref{Framework}.

We can relate the trajectory of the physics sense \(A^{T} = \left\lbrace{A^{1}, A^{2},\ldots{,}A^{l}}\right\rbrace\) in time \(T=\lbrace{t}\rbrace\) with the trajectories of all other physics senses \(\lbrace{B^{t}_{\gamma}}\rbrace\) from the totality \(\Gamma\setminus\left\lbrace{A^{t}}\right\rbrace\). Our goal here is to characterize information transmission between senses' trajectories via their path mutual information
\begin{equation}
\label{t}
\mu\left(A^{T}\right) = \sum_{t\in T}\sum_{\gamma=1}^{\dim(A)}p_{(A^{t},B^{t}_{\gamma})}(\alpha,\beta)\ln\frac{p_{(A^{t},B^{t}_{\gamma})}(\alpha,\beta)}{p_{A^{t}}(\alpha)p_{B^{t}_{\gamma}}(\beta)} ,
\end{equation}
where \(p_{(A,B)}(\cdot)\) is the joint probability mass function of \(A\) and \(B\), \(p_{A}(\cdot)\) and \(p_{B}(\cdot)\) are the marginal probability mass function of \(A\) and \(B\), respectively. The quantity \(\mu\left(A^{T}\right)\) will be called the trajectory mutual information (TMI, from now on) of physics sense \(A\).

To derive a distribution of probabilities \(\mu\left(A^{T}\right)\), we must characterize the principle whereby out of the set of all possible quotient spaces \(\lbrace{\left(\Omega_{A},\varpi_{A}\right)}\rbrace\) (see \ref{Framework}), we select those that are realized. The comparison of quotient topological spaces we have drawn here allows for formulation of an extreme principle for the changeability of a physics sense: from the given state \(\left(\Omega_{A},\varpi_{A}\right)\) the sense tends towards a state \(\left(\Omega_{B},\varpi_{B}\right)\) with the maximum value of \({\mathsf{P}}\lbrace{B}\rbrace\) (Eq.~\eqref{eq:g2}). We now admit that the actual trajectory of the physics sense represents a most probable trajectory. This is equivalent to the transition to a physics sense with the maximum value of the TMI. It is easy to derive the Fr\'{e}chet extreme distribution for TMI.

The challenge facing us at this point is to approximate the probability distribution function of \(\mu\left(A^{T}\right)\). The quantity of TMI represents a sum total of the information activity of a physics sense, understood as a sum of its own activity and that of the senses whose orbits intersect its own. We are dealing here with values of the sums of the form
\begin{equation}
\label{eq:g5}
Z_{\gamma} = \sum_{\gamma=1}^{\dim(A)}\zeta_{\gamma} = \sum_{\gamma=1}^{\dim(A)}\sum_{t\in T}p_{(A^{t},B^{t}_{\gamma})}(\alpha,\beta)\ln\frac{p_{(A^{t},B^{t}_{\gamma})}(\alpha,\beta)}{p_{A^{t}}(\alpha)p_{B^{t}_{\gamma}}(\beta)}, 
\notag
\end{equation}
such that \(\mu\left(A^{T}\right) = Z_{\gamma}\). Since we postulated that the true trajectory of the physics sense \(A\) is corresponding to the maximum of \(\mu\left(A^{T}\right)\), we must consider the limit behavior of extremes
\begin{equation}
\label{eq:g6}
(d\rightarrow\infty)\left(\dim(A)\rightarrow\infty\right)\colon \widehat{Z}_{d\gamma} = \max_{1\leq \delta\leq d}\sum_{\gamma}\zeta_{\delta\gamma}.
\end{equation}
Let the variables \((\delta, \gamma \geq 1)\colon \zeta_{\delta\gamma}\) be mutually independent and have a common distribution \(F\). We assume that \(F\) belongs to the class of subexponential distributions \citep{Foss_Korshunov_2013}: the convolution tail \(\overline{F{\ast}F}(u)\) is equivalent to \(2\overline{F}(u)\) as \(u\to\infty\), where \(\overline{F}(u)= 1 - F(u)\). Assume also that \(F\) belongs to the class of distributions with regularly varying tails, i.e., \(\overline{F}(u) \sim u^{-a}L(u)\), \(u\to\infty\), where \(a>0\) and \(L(u)\) is a slowly varying function \citep{Bingham_Goldie_Teugels_1987}. This is not necessarily true, but it represents a reasonably good approximation in a large number of subsystems of science \citep{ZENG2017}. Under the conditions considered here, can be obtained that the family of extremes similar to \(\widehat{Z}_{d\gamma}\) has well-known Fr\'{e}chet extreme distribution \(\mathsf{P}{\lbrace Z^{\prime} \leq z\rbrace}\rightarrow \exp(-z^{-a})\), \(0<a<2\) \citep{Lebedev_2005_PT}. It means, among other things, that bigger TMI values are normally a result of individual senses having higher information activity rather than a combined activity of a large number of senses.

\subsection{Assistant quantities}
\subsubsection{Similarity}
To derive similarity between \((X_{A},\tau_{A})\) and \((X_{B},\tau_{B})\), one should keep in mind that the value Eq.~\eqref{eq:g2} characterizes the probability of a particular state of a physics sense. It follows that the similarity between physics senses \(A\) and \(B\) will be given by
\begin{equation}
\label{sim_0}
s_{AB} = \ln\frac{p_{(A,B)}(\alpha,\beta)}{p_{A}(\alpha)p_{B}(\beta)} .
\notag
\end{equation}
This expression is nothing but the point-wise mutual information.

\subsubsection{Normalized entropy}
To suppress the dependence of the entropy \(h(A)\) on \(k\) in Eq.~\eqref{eq:g3} and keeping in the mind Eq.~\eqref{eq:dim}, we consider the normalized entropy \(h_{n}(A)\) defined as followed:
\begin{equation}
\label{eq:g4}
h_{n}(A) = \frac{h(A)}{\ln{\dim(A)}}.
\end{equation}

\subsubsection{Complexity}
Following \citep{LOPEZRUIZ_1995, MARTIN2006439, Rosso_2007}, we computed the values of the so-called statistical complexity
\begin{equation}
\label{compl}
c(A) = h_{n}(A)d(A),
\end{equation}
where the ``disequilibrium'' \(d(A)\), which measures the deviation of the probability distribution \(\mathsf{P}\lbrace{A}\rbrace\) from uniform probability distribution, is defined as
\begin{equation}
\label{diseq}
d(A) = \sum_{\alpha=1}^{\dim(A)}\left(p(\alpha) - \frac{1}{\dim(A)}\right)^{2} .
\notag
\end{equation}
By construction, the quantity \(d(A)\) vanishes for probability distributions \(\mathsf{P}\lbrace{A}\rbrace\) that correspond to absolute order and maximal randomness.

\subsubsection{Transitivity}
The fraction of pairs of physics senses that are themselves intersected defines the sense's transitivity \(tr(A)\) \citep{Newman_2018}: 
\begin{equation}
\label{tr}
tr(A) = \frac{\#\: \mathsf{pairs\: of\: senses\: whose\: orbits\: intersect\: with\:} O_{A}}{\#\: \mathsf{pairs\: of\: senses}} .
\end{equation}
The transitivity is an important quantity, which informs on the density of a sense's information field.

\subsubsection{Citation rate}
To construct the citation rate of the concepts, we used a three-year citation window and fractional counting. It means the citation rate of the concept \(x\) of document \(y\) is a share of all other concepts of this document. Thus the total concept citation rate is the sum of all its fractional citations in all documents where it is used.

\section{Data}
The publication data for this study comes from the INSPIRE database. INSPIRE is a leading information system in HEP \citep{Xu_2020_101005} that provides open access to more than \(1\) million scientific publications. It continues the SPIRES physics literature information system. The INSPIRE database is the result of close collaboration between such institutes as CERN, DESY, Fermilab, and SLAC. It brings together information from arXiv.org, NASA--ADS, PDG, and other HEP publishers. In addition to documents, INSPIRE integrates information about authors, institutes, journals, conferences, and jobs in the field of HEP. This complex integrated system provides opportunities to analyze information processes in HEP from different points of view. The INSPIRE system is of interest not only to physicists but also to researchers in quantitative science studies \citep{Zingg_2020}. Researchers who have worked with INSPIRE datasets pointed out that the data is well-curated and easily accessible \citep{jang_how_2019, strumia_biblioranking_2019}.

Another significant aspect of the INSPIRE information system is the keywords that accompany documents. The keywords are automatically extracted with the help of regular expressions from full-text papers by the BibClassify extraction algorithm\footnote{For details, \url{http://cds.cern.ch/help/hacking/bibclassify-extraction-algorithm}, last accessed on 08 December 2020.}. This algorithm draws upon a controlled vocabulary resulting from the HEP Index (HEPI). The HEPI thesaurus is maintained by the German High Energy Physics Laboratory DESY since 1963. It is prepared by the subject specialists and updated occasionally. The HEPI thesaurus is actively used by the principal HEP institutes and information databases. The BibClassify algorithm extracts two types of keywords: single keywords and composite keywords. Composite keywords are combined with several single keywords. If single keywords are placed next to each other, they form a composite keyword. The words of composite keywords are separated by a colon e.g. ``Higgs particle: mass'' or ``CP: violation''. The final keyword list consists of top \(20\) first best single and composite keywords. These keywords serve as an information base for the study.

Our dataset counts documents with the term ``new physics'' in titles, abstracts, or keyword lists published in 1989---2018\footnote{Data was retrieved from the INSPIRE database in October 2019. Processing of the XML dump, data preparation, and statistical analysis were performed with the R programming language. Citation data was downloaded from the INSPIRE database in August 2021.}. To make the sample more stable and consistent, we excluded some irrelevant documents and concepts. First of all, we removed the documents featuring only one physics concept or no concepts at all. Secondly, we removed keywords having errors due to the indexing procedure and keywords, denoting document types like ``book'', ``thesis'', ``report'', and so on. After these procedures, the final dataset counted \(19{,}830\) documents and \(8{,}822\) physics concepts. The collection of documents thusly obtained was subjected to further statistical analysis.

\section{Results and discussion}
\subsection{Statistical features of new physics discourse}
\begin{figure}[ht!]
\centering
  \includegraphics[width=\linewidth]{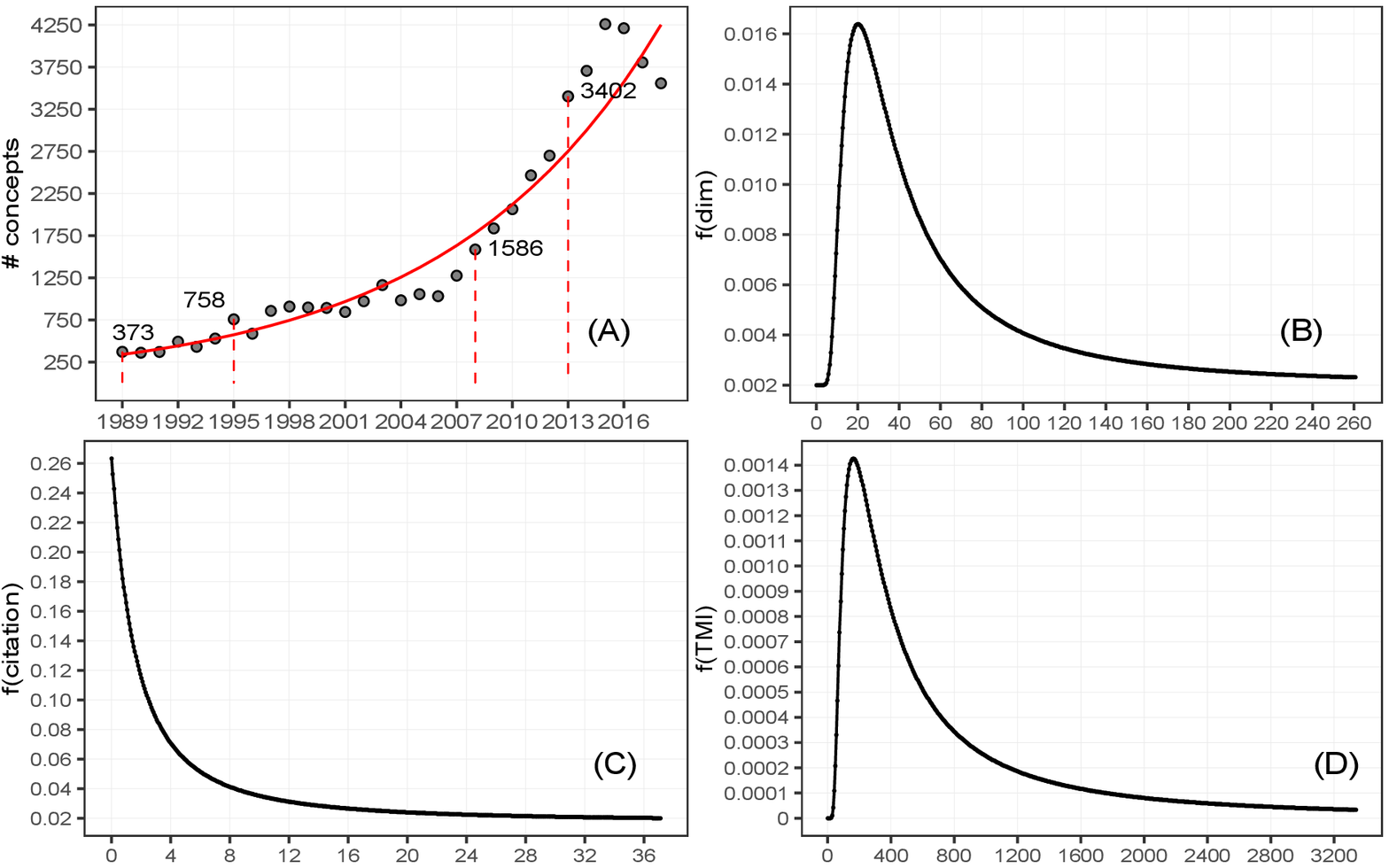}
\caption{Features of new physics discourse.\\
(A)~Dynamics of new physics concepts in 1989---2018. The equation of the regression curve: \(\#\:\mathsf{concepts} = 312.612\cdot\exp(0.087t)\) \(\left(R^{2}=0.954, p=0.000\right)\). Dashed vertical lines indicate the generations of physics senses (see \ref{5.3}~``Generations of physics senses'').\\
(B)~The approximation of the probability density function of the dimension: a Fr\'{e}chet distribution with parameters \(\alpha = 1.1978, \: \beta = 34.672\) (Kolmogorov--Smirnov statistic \(0.0137\).)\\
(C)~The approximation of the probability density function of the citation: a Pareto Type II distribution with parameters \(\alpha = 0.7986, \: \beta = 3.0318\) (Kolmogorov--Smirnov statistic \(0.0145\).)\\
(D)~The approximation of the probability density function of the TMI: a Fr\'{e}chet distribution with parameters \(\alpha = 0.8947, \: \beta = 382.26\) (Kolmogorov--Smirnov statistic \(0.0128\).)}
\label{fig2}      
\end{figure}

Fig.~\ref{fig2}A illustrates the number of new physics unique concepts indexed in INSPIRE in the years 1989---2018. It can be seen that the figures gradually increase from between 1989 and 1995. Then they show some fluctuation to proceed with rapid growth after 2006. In 2006---2015, one can note a quick rise of numbers and even a spike in 2013 and 2015. In 2015, the number of concepts reached its maximum value of \(4259\) items, followed by the curve dropping sharply to \(3557\) concepts in 2018. For the whole 30--year period, the number of unique concepts rises from \(373\) to \(4259\) items, and the curve fits the exponential regression line (cf. \citep{milojevic_quantifying_2015}).

Fig.~\ref{fig2}B, Fig.~\ref{fig2}C, and Fig.~\ref{fig2}D show approximations of the probability density functions of dimension, citation, and TMI, respectively. One can spot at a glance the qualitative difference of the dimension and the TMI probability distributions from that of citation. Indeed, the probability density functions in Fig.~\ref{fig2}B and Fig.~\ref{fig2}D are of Fr\'{e}chet type, whereas Fig.~\ref{fig2}C shows a power-law distribution function. It seems only natural that  the citation distribution can be fitted by a Pareto distribution \citep{ZENG2017, Fortunatoeaao0185}. The Fr\'{e}chet distribution observed for dimension is untypical of scientometric quantities of this kind (cf. \citep{Boerner2019}). A quantity homologous to dimension, the ``node degree'' is normally distributed according to a power law \citep{ZENG2017, Palchykov_2021_1}. We have, in turn, predicted the Fr\'{e}chet distribution for TMI.

Despite the obvious differences between the dimension and the TMI probability distributions, on the one hand, and that of citation, on the other, they share an internal similarity. The fact is, the Pareto distribution belongs to the maximum domain of attraction of the Fr\'{e}chet distribution \citep{Foss_Korshunov_2013}. The asymptotic behavior of citation distribution is similar to that which defines the distribution of dimension and TMI. In particular, larger values of citation, dimension, and TMI are achieved not through a larger number of physics senses but due to individual physics senses with higher activity.

The Fr\'{e}chet extreme distribution is a distribution of maximums; it expresses relatively enduring structures of the dimension and the TMI. These structures are characterized via the probability of extreme events (exceedances of a high level). Both the dimension Eq.~\eqref{eq:dim} and the TMI Eq.~\eqref{t} are high outliers by nature, so it is only logical that they are described by the Fr\'{e}chet extreme distribution. This distribution is a pattern or arrangement of physics senses; we can surmise, therefore, that it is the trends defined by record values that form the basis of a nascent new physics discourse.

The citation structure of physics senses is, in turn, described by the Pareto distribution. This distribution signifies the fact that very few physics senses possess an extremely high citation rate, while very many demonstrate an extremely low one.

\begin{table}[ht!]
\caption{The Top--30 physics senses of the PageRank value}
\footnotesize
\begin{tabular}{|l|l|r|}
\hline
PageRank & Physics sense & \(\dim\) \\ \hline
1 & numerical calculations & 5762 \\
2 & supersymmetry & 3962 \\
3 & cp: violation & 3523 \\
4 & new physics: search for & 3040 \\
5 & p p: scattering & 3007 \\
6 & higgs particle: mass & 2506 \\
7 & electroweak interaction & 2398 \\
8 & numerical calculations: monte carlo & 2351 \\
9 & neutral current: flavor changing & 2324 \\
10 & minimal supersymmetric standard model & 2287 \\
11 & atlas & 2238 \\
12 & electron positron: annihilation & 2119 \\
13 & cms & 2138 \\
14 & quantum chromodynamics & 2052 \\
15 & dark matter & 2099 \\
16 & effective lagrangian & 2013 \\
17 & effective field theory & 2099 \\
18 & neutrino: mass & 2055 \\
19 & p p: colliding beams & 1902 \\
20 & neutrino: oscillation & 1823 \\
21 & higgs particle: coupling & 1781 \\
22 & lepton: flavor: violation & 1725 \\
23 & lhc-b & 1644 \\
24 & coupling: yukawa & 1784 \\
25 & numerical calculations: interpretation of experiments & 1512 \\
26 & angular distribution: asymmetry & 1600 \\
27 & electroweak interaction: symmetry breaking & 1643 \\
28 & interpretation of experiments & 1504 \\
29 & batavia tevatron coll & 1608 \\
30 & top: pair production & 1592 \\ \hline
\end{tabular}
\label{tab1}
\end{table}
\normalsize

\begin{figure}[ht!]
\centering
  \includegraphics[width=\linewidth]{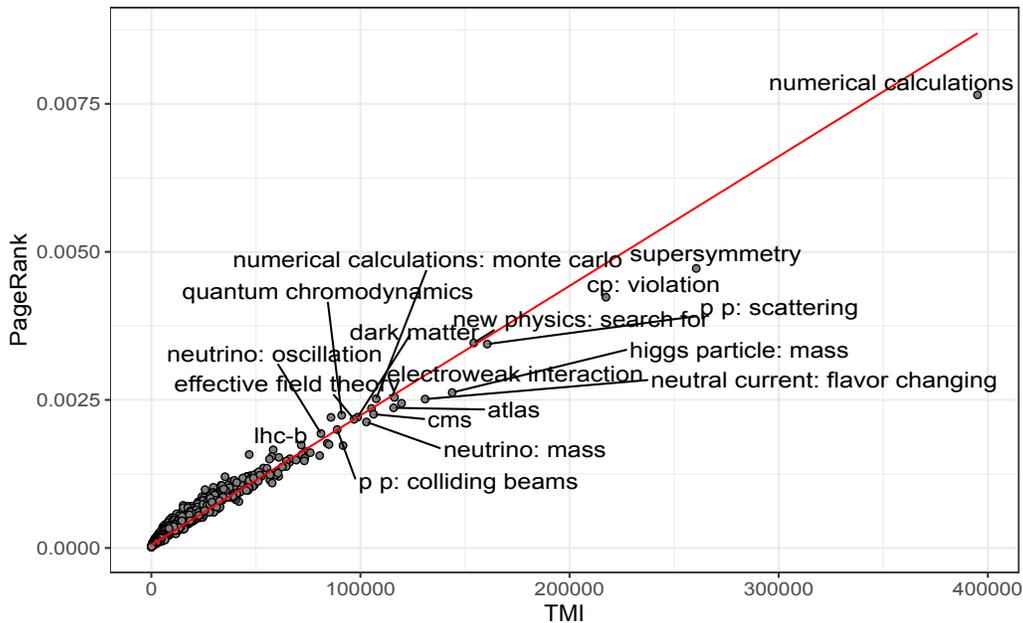}
\caption{Scatterplot of PageRank against the TMI.}
\label{fig8}      
\end{figure}

\begin{figure}[hb!]
\centering
  \includegraphics[width=\linewidth]{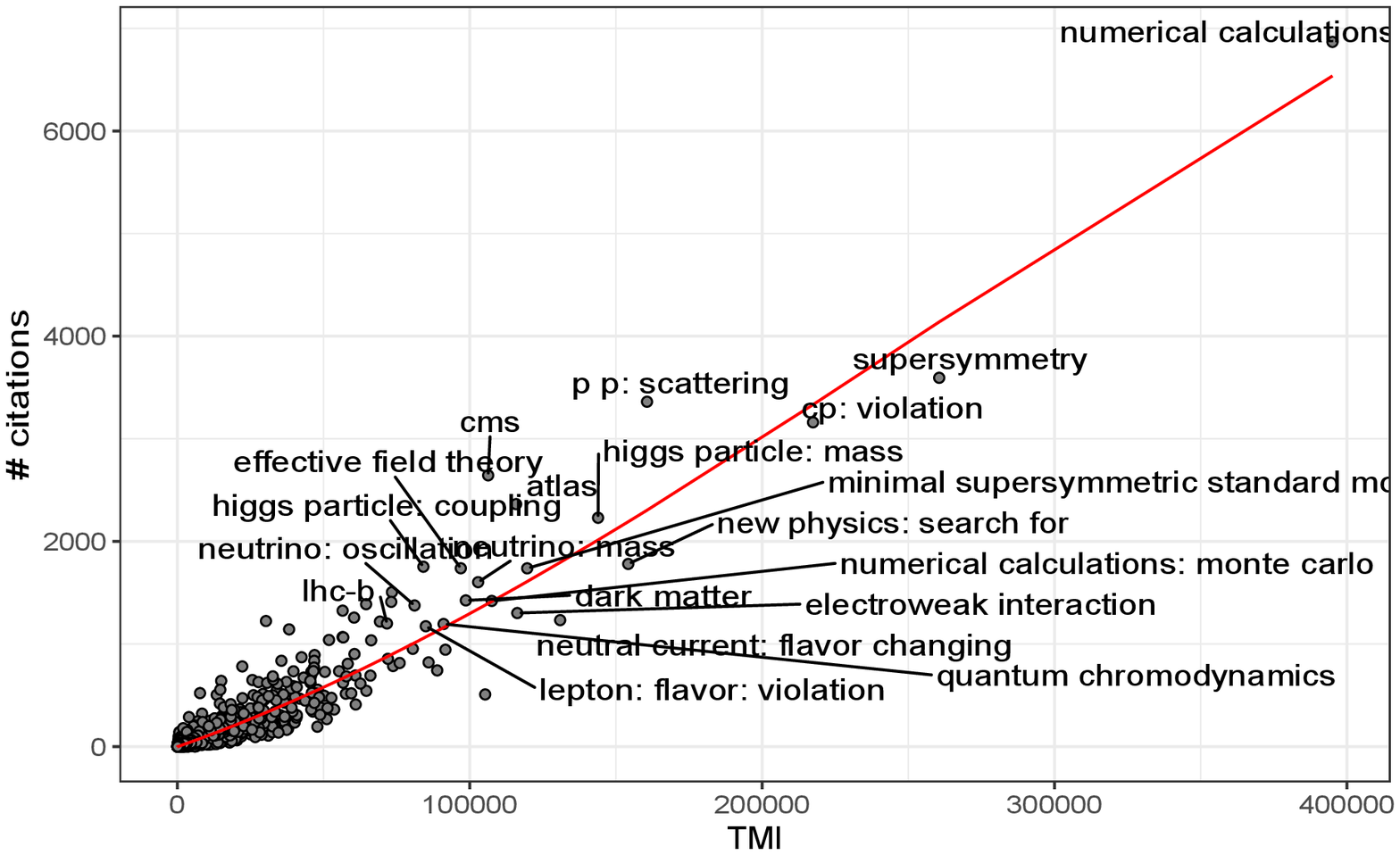}
\caption{Scatterplot of \(\#\: \mathsf{citations}\) against the TMI. The equation of the regression curve of \(\#\:\mathsf{citations}\) is the following:~\(\#\:\mathsf{citations}= 1{.}547\cdot{10}^{-8}\cdot\mu^{2}+0{.}11\cdot\mu - 3{.}278 \: \left(R^{2}=0.903,\:p=0.000\right)\).}
\label{fig3}      
\end{figure}

\subsection{PageRank and TMI}
In order to study the dynamics of the ``importance'' and ``authority'' of physics senses in the years 1989--2018, the PageRank value of physics senses was calculated with the use of the \(\mathsf{igraph}\) software package. The Top--30 physics senses of the PageRank value are listed for illustration and comparison, as shown in Table \ref{tab1}.

Fig.~\ref{fig8} shows PageRank as a linear function of the TMI. The relationship between the two quantities is described by the regression equation \(\mathsf{PageRank} = b\cdot\mu\left(A^{T}\right)+a\), where \(b=2.187\cdot{10^{-8}}\), \(a=5.157\cdot{10^{-5}}\) (\(R^{2}=0.958\), \(p=0.000\)). The importance of physics senses is directly proportional to their measure of the mutual dependence: the stronger the sense is connected with others, the higher its authority. This means that the authority of a sense is determined by its interconnections with a totality of physics senses. This interrelationship between PageRank and the TMI illustrates the structure of the informational process of new physics.

\begin{table}[ht!]
\caption{The characteristics of generations}
\footnotesize
\begin{tabular}{|l|l|l|l|l|}
\hline
Generation & Years & Total number & Number of & Percent of \\
	&	& of senses & new senses & new senses \\
\hline
I & 1989---1994 & 1363 & 1363 & 15.4 \\
II & 1995---2007 & 4071 & 3030 & 34.3 \\
III & 2008---2012 & 4952 & 2274 & 25.8 \\
IV & 2013---2018 & 7183 & 2156 & 24.4 \\
\hline
\end{tabular}
\label{tab2}
\end{table}
\normalsize

\subsection{Generations of physics senses} \label{5.3}
``Generations'' offer a convenient method of classifying physics senses. We chose the doubling of the number of senses in 1995, 2008, and 2013 as the criterion for identifying generations. Following this way, we got four generations (see Table~\ref{tab2} and Fig.~\ref{fig2}A). A physics sense belongs to a particular generation if its first occurrence happens in the period corresponding to that generation. Thus, all senses that occur in the years 1989--1994 are generation I senses. Generation II senses are those that occurred for the first time in documents published in 1995--2007, etc. Table \ref{tab2} shows that the total number of senses increases from generation to generation, although new senses become less common over time. Of the total dataset, \(15.4\%\) are generation I, \(34.3\%\) generation II, \(25.8\%\) generation III, and \(24.4\%\) generation IV senses. Analysis of the generations helped to identify the ``core'' of physics senses---senses that occur in the discourse of new physics throughout all four generations. Such senses amount to \(7.3\%\) of the total dataset. They comprise the most frequently occurring physics senses with connections to many other senses. Their average \(\dim\) is \(471\). For reference, an average \(\dim\) for senses that occur for the duration of three generations is \(204\), whereas senses occurring for the period of one or two generations have an average \(\dim\) of \(34\) and \(74\), respectively. 

\begin{figure}[ht!]
\centering
  \includegraphics[width=\linewidth]{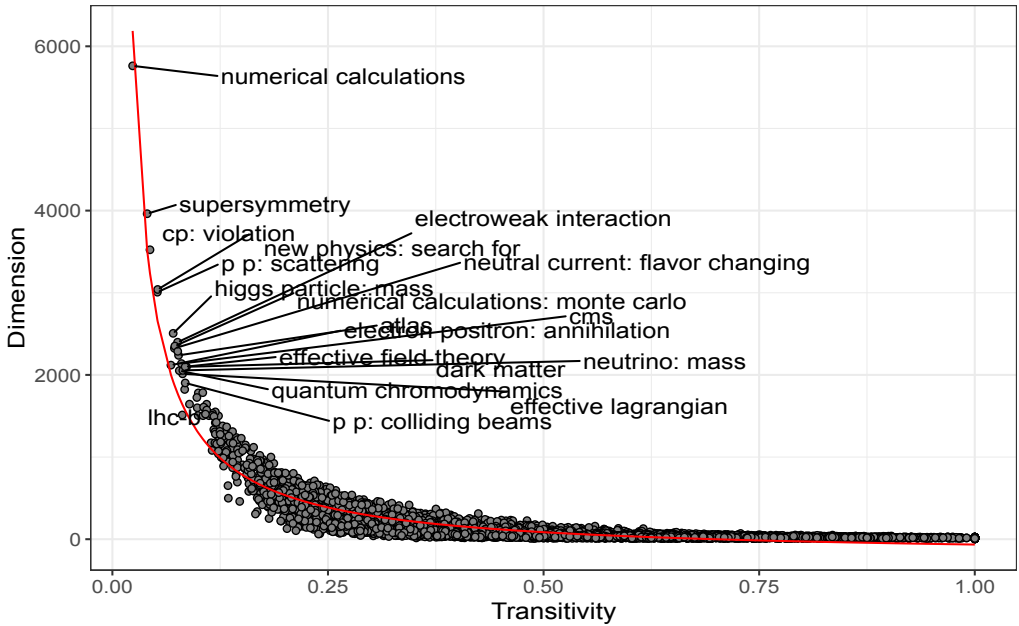}
\caption{Scatterplot of the dimension against the transitivity. The equation of the regression curve of \(\dim\) is the following:~\(\dim = 150{.}418\cdot{tr^{-1}} - 215{.}254\: \left(R^{2}=0.895,\:p=0.000\right)\).}
\label{fig4}      
\end{figure}

\subsection{TMI, dimension, number of citations, and number of documents}
\begin{figure}[hb!]
\centering
  \includegraphics[width=\linewidth]{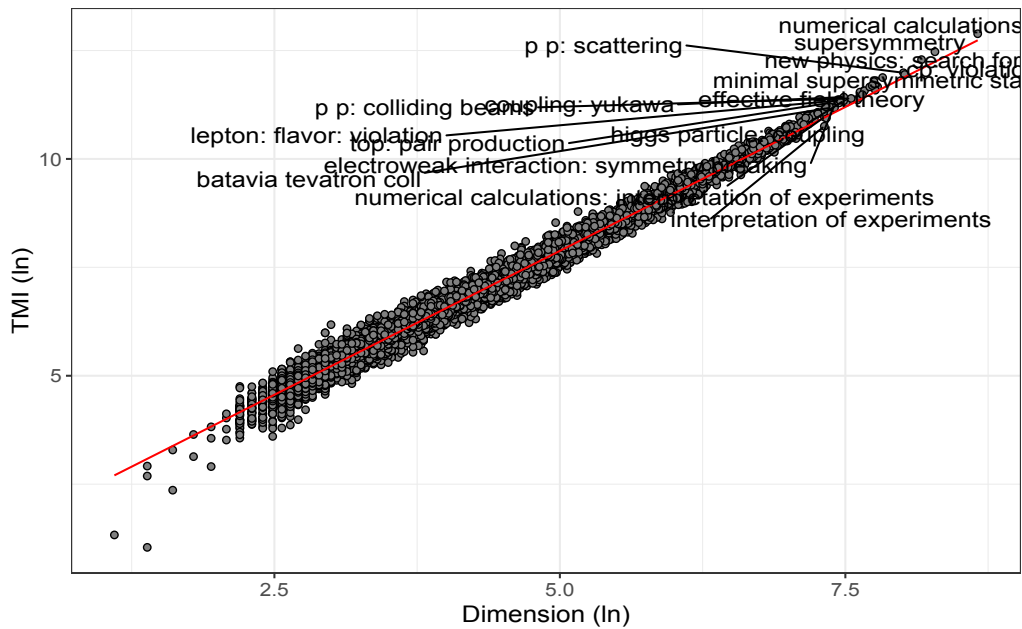}
\caption{Scatterplot of the TMI~(\(\ln\)) against the dimension~(\(\ln\)). The equation of the regression curve of \(\ln\mu\) is the following:~\(\ln\mu = 1{.}327\cdot\ln\dim + 1{.}245\: \left(R^{2}=0.981,\:p=0.000\right)\).}
\label{fig5}      
\end{figure}

\begin{figure}[ht!]
\centering
  \includegraphics[width=\linewidth]{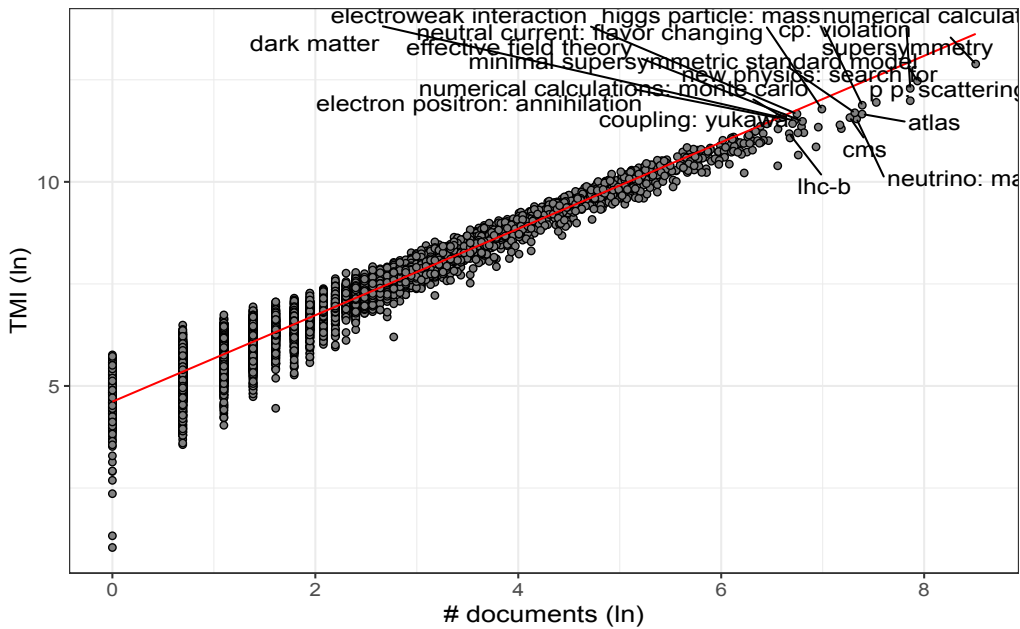}
\caption{Scatterplot of the TMI~(\(\ln\)) against \(\ln\#\:\mathsf{documents}\). The equation of the re\-g\-res\-sion curve of \(\ln\mu\) is the following:~\(\ln\mu = 1{.}059\cdot\ln\#\:\mathsf{documents} + 4{.}615\: \left(R^{2}=0.936,\:p=0.000\right)\).}
\label{fig6}      
\end{figure}

Fig.~\ref{fig3} through Fig.~\ref{fig6} show interconnections between quantities essential for understanding the structure of physics senses. 
\begin{itemize}
\item In Fig.~\ref{fig3}, we see TMI as a quadratic function of the number of citations. This is the way in which the Fr\'{e}chet extreme distribution manifests itself: maximums of different values belong to the same physics senses. Fig.~\ref{fig3} indicates that, for the orbit of the sense, its measure of intersection with other orbits directly depends on the citation rate of the sense. This is how the difference between the Fr\'{e}chet distribution of the TMI and the Pareto distribution of the citation expresses itself: the quadratic dependence stems from the fact that the citation is distributed more unevenly than the TMI.
\item Fig.~\ref{fig4} shows a dimension as an inverse function of transitivity. This is in line with results already published in the literature \citep{Newman_2018}. This can be easily interpreted: the higher the value of the dimension, the bigger the number of pairs of physics senses whose orbits intersect with the orbit of the given sense and, therefore, the lower the transitivity value (see Eq.~\eqref{tr}). Since physics senses with a high dimension value are few, and those with a low one are many (the fact that determines  the Fr\'{e}chet extreme distribution of the dimension), the chart in Fig.~\ref{fig4} resembles a hyperbola.
\item A TMI---\(\dim\) linear graph in double logarithmic coordinates is shown in Fig.~\ref{fig5}; and Fig.~\ref{fig6} has a TMI---\(\#\)~documents linear graph in double logarithmic coordinates. It is worth emphasizing that the straight line in double logarithmic coordinates indicates a power function. The level of TMI changes in proportion to the dimension to the power of \(1.327\) and the number of documents to the power of \(1.059\). This fact indicates heavy tails of TMI distributions. Moreover, the power functions TMI(\(\dim\)) and TMI(\(\#\)~documents) are stable under changes in the scales of TMI, \(\dim\), and number of documents.
\item Fig.~\ref{fig5} shows the influence of the topology of physics senses on their informational value in the new physics subfield. It appears that knowledge structure is a function of the  dimension. To put it simply, the principal topological characteristic of a physics sense determines its position in the information structure of new physics. Notably, the TMI  grows faster than the linear function of the  dimension. It follows that physics senses with high TMI values play a disproportionately more significant role, lending a hierarchical character to the knowledge structure of the new physics subfield.
\item Importantly, extreme values in all four charts belong to the same physics senses as listed in Table~\ref{tab1}. These Top--30 physics senses determine, to a certain extent, the properties of the scholarly discourse of new physics.
\end{itemize}

\begin{figure}[hb!]
\centering
  \includegraphics[width=\linewidth]{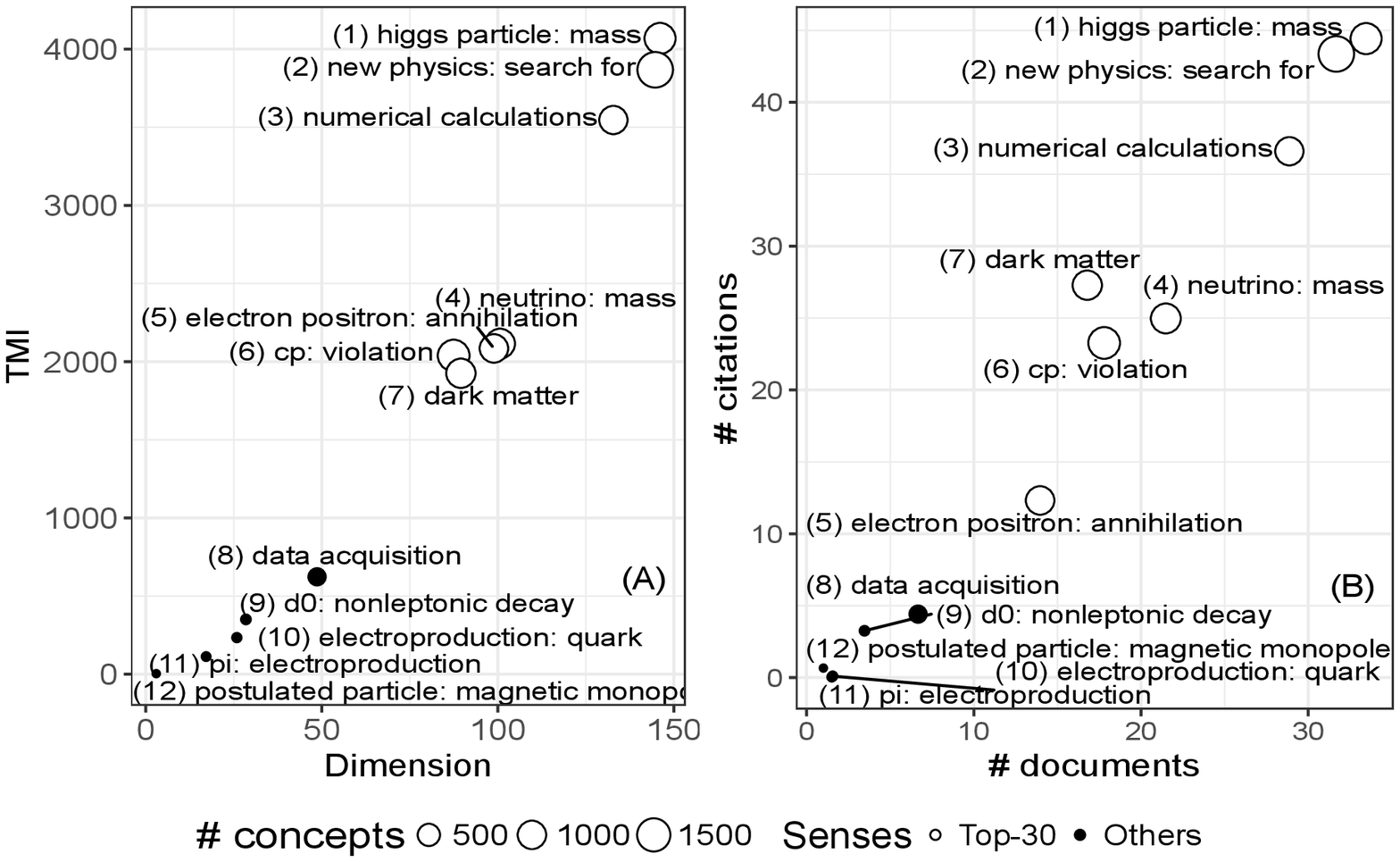}
\caption{Space of physics sense clusters.\\
A.~The clusters of physics senses in the TMI---the dimension plane.\\
B.~The clusters of physics senses in the \(\#\: \mathsf{citations}\)---the \(\#\: \mathsf{documents}\) plane.}
\label{fig7}      
\end{figure}

\begin{figure}[ht!]
\centering
  \includegraphics[width=\linewidth]{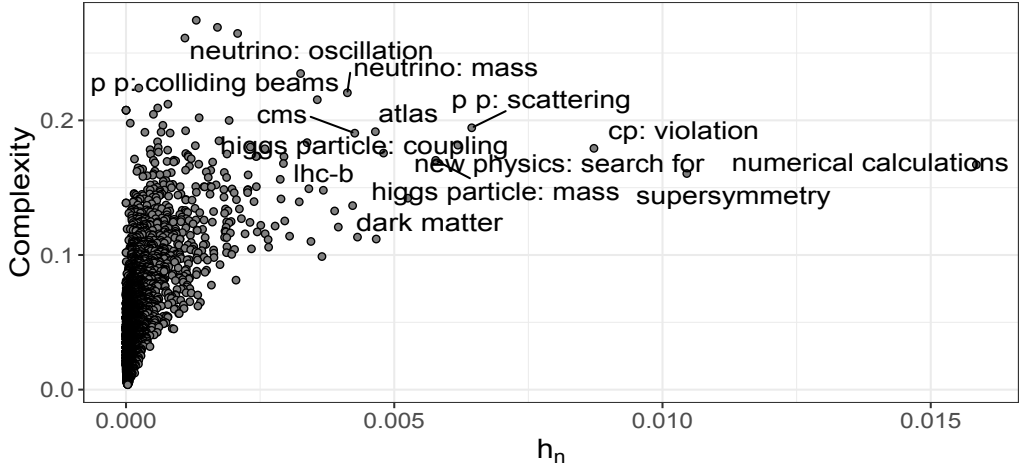}
\caption{Scatterplot of the statistical complexity \(c(A)\) against the normalized entropy \(h_{n}(A)\).}
\label{fig9}      
\end{figure}

\begin{figure}[hb!]
\centering
  \includegraphics[scale=0.75]{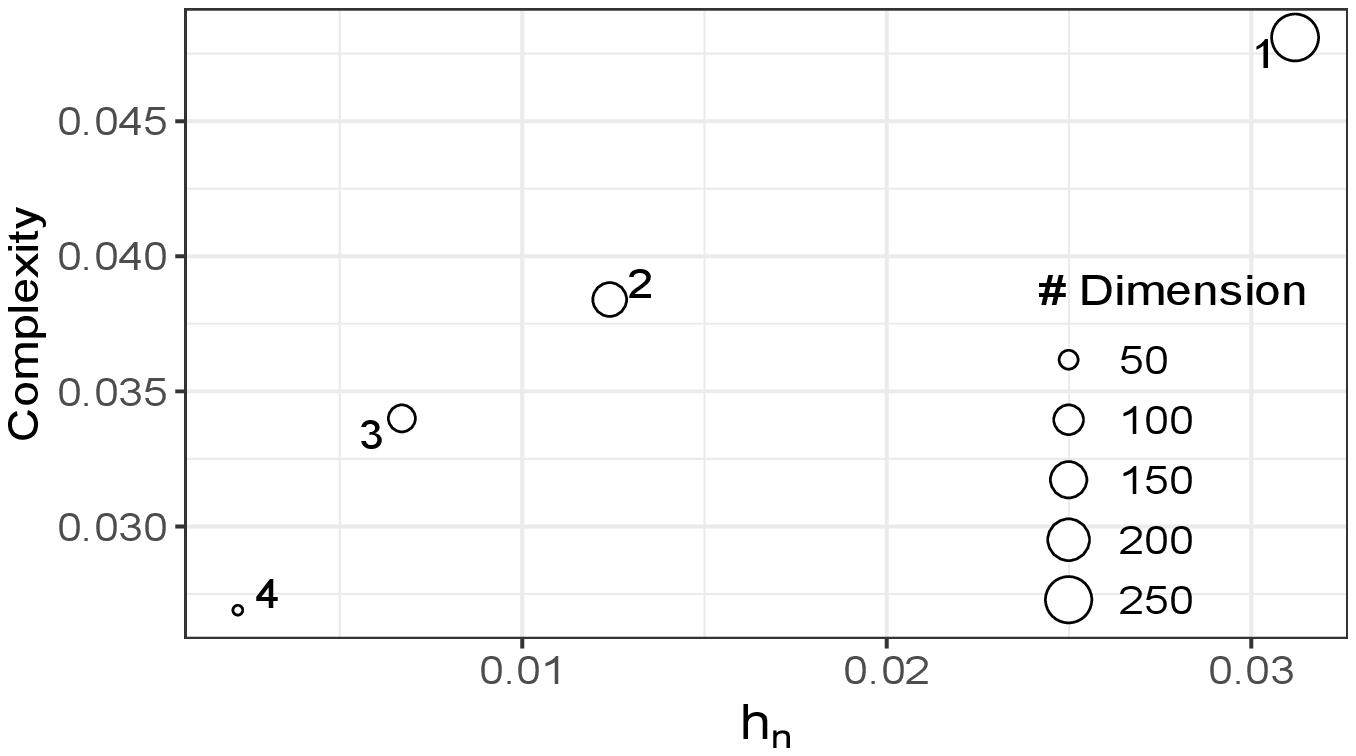}
\caption{The generations of physics senses in the statistical complexity \(c(A)\)---the normalized entropy \(h_{n}(A)\) plane.}
\label{fig10}      
\end{figure}

\subsection{Clusters of physics senses}
We calculated the values of \(s_{AB}\) for all pairs of physics senses and carried out the procedure of sense clustering with the help of the function \(\mathsf{cluster.walktrap}\)\footnote{See \(\mathsf{igraph}\) software package.}. This function uses a random walk algorithm to find dense subgraphs in complex networks (see details in \citep{Masuda_2017_1}). The algorithm assumes that short random walks are in the same community. Based on this, the algorithm tries to find objects that form subgraphs within one network.

Obtained were \(12\) clusters Fig.~\ref{fig7}. We interpret the clusters on the basis of average values calculated for physics senses that form the clusters. Analysis of the behavior of the senses with regard to TMI and \(\dim\) allows us to characterize their information interaction with other senses (see Fig.~\ref{fig7}A). The closer senses get in terms of their information interaction, the bigger their chance of ending up in the same cluster and sharing other similarities. Fig.~\ref{fig7}B shows the interdependence between the citation rate of the senses in a cluster and the number of documents where they occur. The figure shows that the first three clusters are comprised of senses with the highest citation rate and the most frequently occurring in the studied documents. The same three clusters contain senses with the most authority. It is these clusters that boast a concentration of the PageRank's Top--30 senses Table~\ref{tab1}. Of the latter, the majority can be found in the second cluster: it includes ten of the Top--30 senses. For reference, clusters from 8 to 12 do not feature a single sense with a high PageRank value. Clusters 1 through 3 comprise senses with the highest TMI and \(\dim\) values.

As becomes clear from Fig.~\ref{fig7}A and Fig.~\ref{fig7}B, the set  of clusters splits into three groups. The first group contains basic physics senses. Due to their high \(\dim\) and TMI values, they consolidate the scholarly discourse about  new physics. The first group of clusters forms a key part of the HEP subfield that relates to new physics. The second can be interpreted as the mainstream of the discourse; it consists of the main physics senses. Main discourse events can occur within the group's perimeter. Finally, the third group of clusters appears largely unstructured and chaotic. The most likely reason is that it is formed by physics senses that are either peripheral or emerging or declining (cf. \citep{Hosseini_2021}).

\subsection{Complexity and entropy}
The statistical complexity \(c(A)\) and the normalized entropy \(h_{n}(A)\) allow us to study the structural properties of a physics sense. The quantity \(c(A)\) indicates the ``state of disorder'' such that senses with extremely ordered or disordered structures will receive lower scores, while senses with non-trivial structures---higher ones. At the same time, \(h_{n}(A)\) measures the degree of diversity of the support \(X(A)\) of a sense and stochasticity. Thus, the statistical complexity \(c(A)\)---the normalized entropy \(h_{n}(A)\) plane (see Fig.~\ref{fig9}) partially reflects the relationship of the notions of disorder/complexity versus order/simplicity with regard to physics senses. Although these quantities are loosely connected statistically (the Pearson correlation coefficient is equal to \(0.557\) between \(c(A)\) and \(h_{n}(A)\)), their relationship partially encodes structural features of each physics sense. For example, smaller values of \(h_{n}(A)\) and larger values of \(c(A)\) correspond to senses that consist of regular sets of recurring ordinal concepts. By contrast, senses that feature a significant diversity of concepts grouped into stochastic sets will generate larger values of \(h_{n}(A)\) and smaller values of \(c(A)\).

\(75\%\) of physics senses have the value \(c(A)\leq 0.045\) and \(h_{n}(A)\leq 6.67\cdot 10^{-5}\), i.e., they occur in that domain of the statistical complexity \(c(A)\)---the normalized entropy \(h_{n}(A)\) plane that can be identified through the paired notion of order/simplicity. As illustrated in Fig.~\ref{fig9}, the Top--30 physics senses possess the highest values of normalized entropy: from \(0.0159\) for ``numerical calculations'' to \(0,0019\) for ``interpretation of experiments''. However, the values of statistical complexity observed for the leaders of the rating are far from unprecedented: they range from \(0.2348\) (``neutrino oscillation'') to \(0.0988\) (``quantum chromodynamics''). We may argue that physics senses in the Top--30 rated positions are characterized by the highest level of disorder and a fairly high complexity level. It is these structural properties that, on the qualitative level, explain the position of the central senses in the scholarly discourse of new physics.

We studied the evolution of physics senses in the statistical complexity \(c(A)\)---normalized entropy \(h_{n}(A)\) plane. For each physics sense, we calculated the values of \(c(A)\) and \(h_{n}(A)\) by years and then averaged the results by generation. The results are presented for each generation in Fig.~\ref{fig10}. We can see that each successive generation demonstrates lower values of statistical complexity and normalized entropy than their predecessors. This observation conforms to the overall picture as shown in Fig.~\ref{fig9}. As we know, the vast majority of physics senses occur in the domain of low values of \(c(A)\) and \(h_{n}(A)\). At the same time, it follows from Fig.~\ref{fig2}A that the number of physics senses was growing exponentially in time. Therefore, physics senses with lower values of \(c(A)\) and \(h_{n}(A)\) belong mainly to younger generations. In other words, generation I constitutes the semantic core of the scholarly discourse of new physics.

\section{Conclusion}

Our paper contributes to the literature by addressing the question of quantitative research on the new physics knowledge structure.
\begin{enumerate}
\item We develop a new method of keyword co-occurrence analysis that unites physics concepts of different documents and binds different documents with the same physics concepts. It is a more general approach than the network one since it offers a natural framework for quantifying knowledge structure of the new physics subfield. We consider topological structures that reveal relationships among concepts and call them physics senses. These senses represent complex systems of direct and indirect interactions between physics concepts.
\item The proposed phenomenological model has the advantage of providing a relevant quantitative evaluation of the distribution of physics senses without recourse to a relatively limited scientometric framework. In case of the new physics' subfield, the Fr\'{e}chet extreme distribution indicates a high level of competition between physics concepts.
\item Physics senses form certain network clusters. The position of physics sense in the new physics' subfield is completely conditioned by the value of dimension and the belonging to a generation. Our findings indicate that the research interests of physicists are focused on a limited number of physics senses which, due to their higher dimension, act as a kind of glue that holds together a heterogeneous discourse of new physics.
\end{enumerate}

However, some limitations need to be considered. First of all, our results are limited to the content of the INSPIRE database. However, given its state as one of the most comprehensive databases in high energy physics, we believe it provides the most possibly complete picture of new physics discourse. A second potential drawback of our study is that it depends on the BibClassify algorithm that extracts keywords from documents. Nevertheless, BibClassify draws upon the HEP Index maintained by the physics community. This index manifests the state of physics discourse as the physicists see it. Therefore, despite the above shortcomings, we believe our study gives a relevant view of the historical dynamics of new physics discourse.

That various potential studies can be conducted in the future using the proposed methodology. It would be interesting to carry out a comparison of several HEP fields: for example, new physics vs. the standard model. Also, assessment and improvement of the suggested methodology would benefit from the involvement of new data sources and other text analysis tools.

\appendix
\section{Framework}
\label{Framework}
\small
Recall that a partition (or a cluster expansion) of \(X\) is a closed disjoint cover of \(X\), and each partition of a set \(X\) determines the equivalence relation on physics concepts: Two elements of \(X\) are said to be equivalent if they belong to the same cluster of the partition. In what follows, we shall denote by \(\Omega\) a fixed cluster expansion of the finite \(T_{1}\) topological space \((X,\tau)\). There is defined a canonical projection \(\pi\colon X\rightarrow\Omega\) of the discrete topological space \((X,\tau)\) onto the set \(\Omega\), i.e. the mapping \(\pi\) assign to the physics concept \(x\in X\) the equivalence class \([x]\in\Omega\). We can now assign to \(\Omega\) a quotient topology on \(\Omega\) by calling a set \(\omega\subset\Omega\) to open if and only if its preimage \(\pi^{-1}(\omega)\) under the canonical projection is open in \(X\) \citep{Singh_2019}. The collection of these sets \(\lbrace{\omega}\rbrace\) is a topological structure \(\varpi\) in the quotient set \(\Omega\), and \(\Omega = \cup_{\alpha\in X}\omega_{\alpha}\), \(\omega_{\alpha}\cap\omega_{\beta}=\varnothing\) when \(\alpha\notin\beta\). We stress that by construction, each cluster \(\omega_{\alpha}\) of the cluster expansion \(\Omega\) corresponds uniquely to one physics concept \(x_{\alpha}\) according to the configuration \(\sigma\) (see Eq~\eqref{config}), and each cluster \(\omega_{\alpha}\) is indiscrete subspace of \((X,\tau)\).

Here \(\vert\cdot\vert\colon X\rightarrow\mathbb{N}\) denotes the counting measure, so \(\vert{X}\vert = M\) is the number of physics concepts in the support \(X\), \(\vert\omega_{\alpha}\vert= m_{\alpha}\) is the number of elements in the cluster \(\omega_{\alpha}\), and \(\sum_{\alpha}m_{\alpha}={M}\). We can introduce the corresponding non-strict partial order on \((\Omega,\varpi)\): \({\omega_{1}}\preceq{\omega_{2}}\preceq{\ldots}\preceq{\omega_{k}}\). Here, the non-strict partial order \(\preceq\) on the set \(\lbrace m_{\alpha}\rbrace\) gives the non-strict partial order \(\preceq\) on \((\Omega,\varpi)\) --- on the maximal \(T_{0}\)-quotient of \((X,\tau)\). Hence, a finite topological space \((X,\tau)\) is fully described by a non-strict partially ordered set \(\left(\Omega,\preceq\right)\) whose clusters \(\omega_{1},\omega_{2},\ldots,\omega_{k}\) are equipped with multiplicities \(m_{1},m_{2},\ldots,m_{k}\).

The nonempty set \(C_{\Omega}(\omega_{\alpha}) = \lbrace\omega_{\alpha}\in \Omega\colon\omega_{\alpha}\preceq\omega_{\beta}\rbrace\) is said to be a cone of \(\omega_{\alpha}\) over \(\left(\Omega, \preceq\right)\). We can look at this in another way: If \(\omega_{\beta}\in C_{\Omega}(\omega_{\alpha})\), then \(\omega_{\alpha}\) precedes or equal to \(\omega_{\beta}\), i.e. \(\omega_{\alpha}\preceq\omega_{\beta}\).  For elements \(\omega_{\alpha}\), \(\omega_{\beta}\) of \(\left(\Omega, \preceq\right)\), \(C_{\Omega}(\omega_{\alpha}) = C_{\Omega}(\omega_{\beta})\) implies \(\omega_{\alpha} =\omega_{\beta}\), and \(C_{\Omega}(\omega_{1})\subset C_{\Omega}(\omega_{2})\subset\cdots\subset C_{\Omega}(\omega_{k})\) implies \(\omega_{1} \preceq\omega_{2} \preceq {\ldots} \preceq\omega_{k}\).

The order topologies on the quotient space \((\Omega,\varpi)\) are in bijective correspondence with the relation \(\preceq\). The collection of all cones \(C_{\Omega}(\omega_{\alpha})\) forms a minimal base for the topology \(\varpi\) \citep{Singh_2019}. Further, the dimension \(\dim(A)\) is the maximum of all non-negative integers \(k\) such that exists a chain \(\omega_{1}\preceq\omega_{2}\preceq{\ldots}\preceq\omega_{k}\) in \(\Omega\):
\begin{equation}
\label{eq:dim}
(\omega_{\alpha}\in \Omega)(k \in \mathbb{N}) \colon \dim(A) = \max\left\lbrace{k}\colon \exists\: {\omega_{1}\preceq\omega_{2}\preceq{\ldots}\preceq\omega_{k}}\right\rbrace.
\end{equation}
Since our key intuition is to treat the order topology \(\varpi\) as an internal property of \(\Omega\), our next task is to find an easy way of computing some topological quantity that characterizes \((\Omega,\varpi)\). We can describe the difference between the physics sense \(A\) and physics sense \(B\) by mapping quotient space \(\left(\Omega_{A},\varpi_{A}\right)\) into quotient space \(\left(\Omega_{B},\varpi_{B}\right)\). The topological structures on \(\Omega_{A}\) and \(\Omega_{B}\) are completely identical if there exists a homeomorphism of \(\left(\Omega_{A},\varpi_{A}\right)\) onto \(\left(\Omega_{B},\varpi_{B}\right)\) (in topological terms, these spaces are homeomorphic: \(\left(\Omega_{A},\varpi_{A}\right) \cong \left(\Omega_{B},\varpi_{B}\right)\)). Let us remember that a homeomorphism is an invertible continuous map \(\varphi\colon \left(\Omega_{A},\varpi_{A}\right)\to \left(\Omega_{B},\varpi_{B}\right)\) such that its inverse is continuous \citep{Singh_2019}. Physics senses \(A\) and \(B\) are homeomorphic if and only if there exists a continuous bijective map between \(\left(\Omega_{A},\varpi_{A}\right)\) and \(\left(\Omega_{B},\varpi_{B}\right)\) that preserves the multiplicities \(m_{1},m_{2},\ldots,m_{k}\).

To give a precise meaning to the statements about distinctions or similarities of physics senses, we need to investigate the mappings of \(\varphi\colon \left(\Omega_{A},\varpi_{A}\right)\to \left(\Omega_{B},\varpi_{B}\right)\). It is clear that, if \(\varphi\) is the homeomorphism, physics senses \(A\) and \(B\) are equivalent from topological point of view. Let \(\nu = \left|\lbrace\varphi\rbrace\right|\) be the cardinality of the set of homeomorphisms. Suppose the equality \(\nu_{A}=\nu_{B}\) holds, then \(\left(\Omega_{A},\varpi_{A}\right) \cong \left(\Omega_{B},\varpi_{B}\right)\). In this case, \(\varphi\) is the a one-to-one correspondence between  each class \(\omega_{\alpha}\in\Omega\) and itself, and we get
\begin{equation}
\label{eq:g1}
\left(\sum_{\alpha = 1}^{\dim(A)} m_{\alpha} = M\right)\colon \nu\left\lbrace\left(\Omega_{A},\varpi_{A}\right)\right\rbrace = \prod_{\alpha = 1}^{\dim(A)}m_{\alpha}^{m_{\alpha}} .
\end{equation}
Let \({\mathsf{P}}\left\lbrace{A}\right\rbrace\) be the ratio of the number of homeomorphisms \(\nu_{A}\) to the number of all possible mappings from \(\Omega\) to \(\Omega\)  
\begin{equation}
\label{eq:g2}
{\mathsf{P}}\left\lbrace A\right\rbrace = \frac{\nu_{A}}{M^{M}}.
\end{equation}
Thus, the number \({\mathsf{P}}\left\lbrace A\right\rbrace\) is the probability of a partition \(\Omega_{A}\) of a support \(X_{A}\) with a fixed topological structure \(\varpi_{A}\). It is obvious that \({\mathsf{P}}\left\lbrace A\right\rbrace\) is proportional to \(M\). We now introduce a normalized logarithmic representation of \({\mathsf{P}}\left\lbrace A\right\rbrace\)
\begin{equation}
\label{eq:g3}
h(A) = - \frac{1}{M}\ln{\frac{\nu_{A}}{M^{M}}} = - \left(\sum_{\alpha=1}^{\dim(A)}\frac{m_{\alpha}}{M}\ln\frac{m_{\alpha}}{M}\right).
\end{equation}
It is readily seen that the right-hand side \(\eqref{eq:g3}\) is nothing other than the entropy, as defined by Shannon.
\normalsize

\section*{Funding}
The article was prepared in the framework of a research grant funded by the Ministry of Science and Higher Education of the Russian Federation (grant ID: 075-15-2020-928).

\bibliographystyle{elsarticle-harv}

\end{document}